\documentclass{article}
\usepackage{graphicx} 
\usepackage{tcolorbox}
\usepackage{algorithmic}
\usepackage[ruled,vlined]{algorithm2e}
\usepackage{booktabs}
\usepackage{amsmath}
\usepackage{mathtools}
\usepackage{amssymb}
\usepackage{tabularx}
\usepackage{url}
\usepackage{amsthm}
\usepackage{hyperref}

\newtheorem{definition}{Definition}

\usepackage[
    top=1in,
    bottom=1in,
    left=1in,
    right=1in,
    includehead,
    includefoot
]{geometry}

\title{Minimum Weighted Feedback Arc Sets \\ for Ranking from Pairwise Comparisons}
\author{Soroush Vahidi, Ioannis Koutis \\
New Jersey Institute of Technology \\ Newark, NJ, USA \\
\texttt{sv96@njit.edu, ikoutis@njit.edu}}
\date{}

\begin{document}

\maketitle

\begin{abstract}
The problem of deriving global rankings from pairwise comparisons has received substantial attention in the literature. Recent work by He et al.~(ICML 2022) introduced learning-based approaches to this task, reporting improvements over existing methods. However, their experimental evaluation did not examine the relationship between ranking from pairwise comparisons and the closely related Minimum Weighted Feedback Arc Set (MWFAS) problem. In this paper, we demonstrate that simple, learning-free combinatorial algorithms for MWFAS can substantially outperform these learning-based approaches, achieving faster runtimes and higher ranking quality on the metrics and benchmark datasets established by He et al.~(ICML 2022).\footnote{This is a preliminary report.}
\end{abstract}

\section{Introduction}
The ranking problem and the Minimum Weighted Feedback Arc Set (MWFAS) problem are closely related; however, the nature of this relationship has not been extensively explored in the literature. Prior work on ranking indicates that only a few studies, such as \cite{FF14} and \cite{C15}, briefly reference the feedback arc set problem and the existence of a polynomial-time approximation scheme (PTAS) for tournaments \cite{KCW07}, without providing a detailed investigation of the connection to MWFAS. To better contextualize this relationship, we first present formal definitions of both problems:

\begin{definition}[The Ranking Problem]
Let $G = (V, E)$ be a directed graph, where $V$ denotes the set of items to be ranked and $E$ is the set of directed edges. Each edge $(i,j) \in E$ has a weight $w_{ij} > 0$, representing the strength of preference for $i$ over $j$ in a pairwise comparison.

A ranking is a mapping $R: V \rightarrow \mathbb{N}$ that assigns an integer rank $R(i)$ to each $i \in V$. The objective is to minimize the total inconsistency with the pairwise comparisons represented by $E$. In particular, we aim to minimize the number (or total weight) of edges for which $R(i) > R(j)$ despite $(i,j)$ indicating a preference for $i$ over $j$.
\end{definition}

\noindent\textbf{Definition (Minimum Weighted Feedback Arc Set Problem).}
Let $G = (V,E)$ be a directed graph, where $V$ is the set of vertices and $E$ is the set of directed edges. Each edge $(i,j) \in E$ has an associated weight $w_{ij} > 0$, representing the cost of removing that edge from the graph.

The objective is to find a subset of edges $E' \subseteq E$ such that removing all edges in $E'$ yields a directed acyclic graph (DAG). The problem seeks to minimize the total weight of the removed edges, that is,
\[
\sum_{(i,j) \in E'} w_{ij},
\]
thereby producing an acyclic graph with the smallest possible removal cost.
\vspace{0.2cm}

The relationship between the ranking problem and the MWFAS problem is rooted in their shared objective of resolving inconsistencies in directed graphs to obtain a meaningful ordering of vertices. In the ranking problem, the goal is to produce an ordering that is as consistent as possible with the given pairwise comparisons, minimizing instances where the assigned ranks contradict the observed preferences. Similarly, the MWFAS problem aims to eliminate inconsistencies by removing a minimum-weight set of edges that break all directed cycles, thereby transforming the graph into a directed acyclic graph (DAG) from which a consistent ordering can be derived.

Once the graph is made acyclic, a topological sort can be applied to obtain a ranking that is consistent with the remaining pairwise comparisons. Consequently, the ranking problem can be viewed as an instance of the MWFAS problem, in which cycles are removed to enable a consistent ordering that reflects the original preferences as closely as possible. It is important to note that the Minimum Feedback Arc Set problem, even in its unweighted form, is NP-hard \cite{K72}, underscoring the inherent computational difficulty associated with deriving consistent rankings in general directed graphs.

\subsection{Ranking Problem and Minimum Weighted Feedback Arc Set (MWFAS)}
The ranking problem can be naturally modeled as an instance of the Minimum Weighted Feedback Arc Set (MWFAS) problem. In this formulation, a directed edge from vertex $i$ to vertex $j$ with weight $w$ indicates that item $i$ is preferred over item $j$ by a magnitude of $w$. If the comparison graph is already a directed acyclic graph (DAG), a topological sort directly yields a ranking in which each edge $(i,j)$ guarantees that $i$ is placed ahead of $j$ in the resulting order.

In practice, comparison graphs often contain cycles, resulting in inconsistencies among the pairwise preferences. To resolve these inconsistencies, the objective is to remove a subset of edges with minimum total weight such that the remaining graph becomes acyclic. This task corresponds directly to the MWFAS problem, which seeks to eliminate a minimal-weight set of edges to obtain a directed acyclic graph (DAG), from which a consistent ranking can then be derived.

The MWFAS problem, like its unweighted counterpart, is known to be computationally challenging, being NP-hard. While the Minimum Feedback Arc Set (MFAS) problem has been extensively investigated in its unweighted form \cite{SS16, KI19, FPP20}, the weighted variant has received comparatively less attention in the literature. The following sections review notable contributions related to the ranking problem.

\subsection{Related Works}
\subsubsection{Previous Work About Minimum Weighted Feedback Arc Set}
\begin{itemize}
    \item The Minimum Weighted Feedback Arc Set (MWFAS) problem has received considerable attention, particularly in the context of approximation algorithms. Karp \cite{K72} introduced the problem and established its NP-hardness. Early work by Even et al.~\cite{EG95} provided an approximation algorithm for MWFAS, which was later improved by Even, Naor, and Rao \cite{EG2000} through a divide-and-conquer approach based on spreading metrics. Although effective, these algorithms rely on the ellipsoid method for solving linear programming relaxations, which makes them computationally expensive in practice.

    Later, Chakrabarti and Imamura \cite{CI03} proposed an $O(VE)$ heuristic based on the local ratio method \cite{BK04}. This approach was shown to be practical and efficient, achieving a $\lambda$-approximation, where $\lambda$ denotes the length of the longest cycle in the graph.

    Baharev et al.~\cite{BA21} presented an exact method for solving the Minimum Feedback Arc Set (MFAS) problem, with a particular focus on sparse directed graphs. Their approach uses lazy cycle enumeration to iteratively extend an incomplete cycle matrix until an optimal solution is obtained. Due to the NP-hard nature of the problem, the algorithm exhibits exponential runtime and does not scale efficiently to large graphs.

\item Hecht et al.~\cite{HM21} proposed an $O(VE^4)$ algorithm for MWFAS. However, the runtime grows rapidly as the number of edges increases, making the method impractical for large or dense graphs.

\item Raman and Saurabh \cite{RS06} investigated the parameterized complexity of the feedback vertex and feedback arc set problems in tournaments, considering both weighted and unweighted variants. They presented fixed-parameter tractable (FPT) algorithms, including an $O((2.4143)^k n^\omega)$ algorithm for the weighted feedback vertex set problem and an $O((2.415)^k n^\omega)$ algorithm for the weighted feedback arc set problem, where $\omega$ is the exponent of matrix multiplication and $k$ is the size of the feedback set. The authors also introduced algorithms for the parametric duals of these problems and examined their complexity in oriented tournaments and general directed graphs.

\item Hassin and Rubinstein \cite{HR94} studied the problem of identifying the largest subset of edges in a directed graph such that the induced subgraph is acyclic. They proposed randomized and deterministic approximation algorithms that achieve improved bounds over previous results, particularly for graphs without two-cycles and for low-degree graphs. Their work also introduced techniques for handling two-cycles optimally and established tight bounds for large classes of directed graphs.

\item Shea-Blymyer and Young \cite{SJ21} investigated the use of satisfiability-modulo-theories (SMT) solvers for solving the Minimum Feedback Arc Set (MFAS) problem. They developed an SMT encoding that exploits incremental solving features, such as unsatisfiable cores. Although promising, their approach remains computationally infeasible for large instances, as solving times scale poorly. The authors also established complexity bounds, demonstrating that MFAS remains challenging even with advanced SMT-based techniques.

\end{itemize}

\subsubsection{Previous Work on the Ranking Problem}
The ranking problem has also been extensively investigated in the literature. Hochberg and Rabinovitch \cite{H00} introduced a ranking method based on pairwise comparisons that aims to maximize the probability of full agreement between a ranking and all corresponding pairwise relationships. Their method was evaluated through portfolio analysis and compared with previous approaches, demonstrating improved prediction accuracy and identifying opportunities for further research.

Cucuringu \cite{C15} proposed Sync-Rank, a method that employs eigenvector and semidefinite programming (SDP) synchronization to address ranking, constrained ranking, and rank aggregation tasks. The approach leverages spectral or SDP relaxations, followed by rounding procedures to recover rankings. Sync-Rank outperformed several competing methods on both synthetic and real-world datasets, while also supporting semi-supervised ranking and identifying consistent partial rankings.

Chen, Gao, and Zhang \cite{CGZ21} studied the ranking of items from incomplete pairwise comparison data within the Bradley--Terry--Luce model. They established the minimax rate for this task with respect to Kendall's tau distance, identifying a phase transition between exponential and polynomial rates depending on the signal-to-noise ratio. The authors proposed a divide-and-conquer algorithm that achieves the minimax rate, demonstrating both theoretical optimality and computational efficiency.

Shah and Wainwright \cite{SW16} introduced a simple, robust, and statistically optimal ranking algorithm based on the Copeland counting method for pairwise comparisons. The algorithm ranks items by counting wins in the comparison graph, offering substantial computational speed-ups over previous approaches. Unlike parametric models such as Bradley--Terry--Luce, the Copeland method does not rely on distributional assumptions and is robust to heterogeneous data types.

Park and Ahn \cite{PA23} introduced the Distinguishing Feature (DF) model, which learns from pairwise comparisons even in the presence of intransitive preferences. Their DF-Learn algorithm, based on a Siamese neural network architecture, outperformed standard models on both synthetic and real-world datasets while capturing complex cyclic dependencies among items.

Gong and Rasingh \cite{GR23} proposed a graph neural network (GNN)–based pairwise surrogate model to accelerate evolutionary optimization by predicting the superiority of candidate solutions rather than directly evaluating their fitness values. Their approach substantially reduced the number of required fitness evaluations while maintaining comparable optimization performance, and is compatible with a wide range of comparison-based evolutionary algorithms, providing a computationally efficient alternative for expensive optimization tasks.

Fang and Bazaza \cite{FB24} introduced Bias-Aware Ranker from Pairwise Comparisons (BARP), an extension of the Bradley--Terry model that incorporates bias parameters to correct evaluator biases and produce unbiased rankings. BARP outperformed several baseline methods on both synthetic and real-world datasets.

Feng and Guo \cite{FG24} introduced Graph Neural Re-Ranking (GNRR), a pipeline that integrates graph neural networks (GNNs) into the re-ranking stage of information retrieval systems. By constructing corpus graphs to model semantic and lexical document relationships, GNRR captures cross-document interactions and improves ranking performance on standard evaluation metrics. Experimental results show consistent performance gains, particularly on challenging datasets, highlighting the potential of GNNs to refine document rankings effectively.

The work most closely related to ours is by Chen and Ghosh \cite{CG21}, who studied vertex ordering in directed graph streams. They addressed the Feedback Arc Set problem by removing feedback arcs to enable topological sorting. However, their focus is on unweighted tournament graphs and lacks empirical validation, whereas we address weighted general graphs and provide extensive experimental evaluation.

Alon \cite{A08} addressed the feedback arc set problem and rank aggregation primarily in tournaments, a restricted class of directed graphs in which every pair of vertices is connected by a directed edge. In contrast, our work generalizes the problem to arbitrary directed graphs, significantly broadening its applicability. Moreover, while Alon's study focuses on theoretical approximation algorithms without implementation or empirical validation, we provide practical heuristics accompanied by implementations and extensive experiments, demonstrating the effectiveness of our approach on real-world datasets.

Bodlaender and Fomin \cite{BF12} presented exact algorithms for several vertex ordering problems on graphs. Their methods run in either $O^*(2^n)$ time and space or $O^*(4^n)$ time and polynomial space, leveraging dynamic programming and recursive techniques akin to the Held--Karp and Gurevich--Shelah approaches. These algorithms generalize techniques used for the Traveling Salesman Problem and can be applied to a variety of vertex ordering tasks, including treewidth and pathwidth. The key contribution of their work is the development of algorithms with moderately exponential complexity applicable to a broad class of graph ordering problems.

Chen \cite{C23} studied the problem of fair rank aggregation by introducing algorithms that enforce fairness across diverse groups in ranked outputs. The author proposed both exact and approximation algorithms for the Closest Fair Ranking and Fair Rank Aggregation problems under several fairness notions, leveraging distance measures such as Kendall’s tau and Ulam distance to obtain efficient and principled solutions.

\section{A Framework for Ranking from Pairwise Comparisons via MWFAS}
Based on our review of existing algorithms, we adopt the heuristic proposed by Chakrabarti and Imamura \cite{CI03} as the core method for our ranking framework, owing to its practical efficiency. The local-ratio based heuristic provides a favorable balance between solution quality and runtime, particularly for graphs with a large number of vertices and edges.

\subsection{Highlights of Our Approach}
\begin{itemize}
    \item \textbf{Speed:} Our method is highly efficient in practice, typically producing results within seconds on the benchmark datasets used in our experiments. In contrast, He et al.~\cite{he22} reported extensive performance tables but did not include runtime measurements, noting only that experiments exceeding one week were marked as ``NaN.'' Other recent works, such as \cite{fogel16} and \cite{NS17}, also did not report computational runtime. To the best of our knowledge, our approach is among the fastest available for this task, while achieving competitive ranking performance.

    \item \textbf{Flexibility:} Our framework treats the MWFAS solver as a modular component, allowing users to select any algorithmic strategy to handle the feedback arc set problem. This design enables flexible control over the trade-off between computational efficiency and solution accuracy: more precise algorithms may be employed when quality is paramount, whereas faster heuristics can be used when scalability is the primary concern.

    \item \textbf{Simplicity:} Our method is conceptually simple and relies on well-established graph algorithms, including topological sorting and cycle detection via depth-first (or breadth-first) search, along with standard heap-based data structures. This simplicity contributes to ease of implementation and facilitates extensibility.
\end{itemize}

\subsection{Algorithm}
\begin{algorithm}[H]
\caption{Ranking from Pairwise Comparisons via MWFAS}
\label{alg:ranking}
\KwIn{A weighted directed graph representing pairwise comparisons.}
\KwOut{A ranking of the vertices.}

\textbf{Step 1: Read Graph and Initialize Data Structures}\;
Parse the input file to construct the set of weighted directed edges $\{(u,v,w)\}$\;
Initialize an adjacency list representation of the graph\;
Store edge weights in a suitable data structure\;

\BlankLine
\textbf{Step 2: Compute a Minimum Weighted Feedback Arc Set (MWFAS)}\;
\While{the graph contains a directed cycle}{
    Identify a directed cycle using depth-first search\;
    Let $\Delta$ be the minimum weight among edges in the cycle\;
    \ForEach{edge $(u,v)$ in the cycle}{
        Decrease $w_{uv}$ by $\Delta$\;
        \If{$w_{uv} = 0$}{
            Remove $(u,v)$ from the graph\;
        }
    }
}
Store all removed edges\;
Sort removed edges in decreasing order of their weights\;
\ForEach{edge $(u,v)$ in the sorted list}{
    \If{adding $(u,v)$ back does not create a directed cycle}{
        Reinsert $(u,v)$ into the graph\;
    }
}

\BlankLine
\textbf{Step 3: Compute Vertex Rankings}\;
Perform a topological sort of the resulting directed acyclic graph\;
Assign rankings according to the topological order\;
Break ties using the score
\[
\frac{\sum_{(i,j)\in E} w_{ij} - \sum_{(j,i)\in E} w_{ji}}
     {\deg^+(i) + \deg^-(i)},
\]
where $\deg^+(i)$ and $\deg^-(i)$ denote the out-degree and in-degree of $i$, respectively\;

\BlankLine
\Return the final ranking and relevant metrics (e.g., removed edges and their weights)\;

\end{algorithm}

\autoref{alg:ranking} implements a straightforward procedure for solving the ranking problem using the Minimum Weighted Feedback Arc Set (MWFAS) framework. The input graph is represented using an adjacency list along with a weight mapping, enabling efficient traversal and edge-weight updates. The algorithm begins by reading the input graph and iteratively removing cycles via depth-first search (DFS), ensuring that the resulting structure becomes a directed acyclic graph (DAG). DFS is also used to evaluate whether removed edges can be reinserted without reintroducing cycles, thereby maintaining connectivity whenever possible. A heap-based topological sort is employed to generate a ranking of the vertices, with ties resolved using a weight-based scoring function. By combining DFS with heap data structures, the algorithm remains both conceptually simple and computationally efficient, offering a practical approach for handling large graphs while minimizing overhead.

\SetAlgoNlRelativeSize{-1} 

\section{Experimental Results}
The tables below show the results of the experiments. The datasets are some of the datasets used in \cite{he22}. The experiments are done on a Macbook Pro with processor 3.3 GHz Dual-Core Intel Core i7 and operating system macOS Monterey version 12.7.6. The code is implemented in Python 3.12.4.

\begin{table}[ht]
\centering
\scriptsize 
\setlength{\tabcolsep}{3pt} 
\renewcommand{\arraystretch}{0.9} %
\resizebox{\linewidth}{!}{%
\begin{tabular}{lcccccccc}
\toprule

\textbf{Dataset Name} & \textbf{Naive Loss} & \textbf{Simple Loss} & \textbf{Ratio Loss} & \textbf{Time (s)} & \textbf{GNNRank Naive Loss} & \textbf{GNNRank Simple Loss} & \textbf{GNNRank Ratio Loss} \\
\midrule
England\_2009\_2010 & 0.13 & 0.53 & 0.60 & 0.01 & 0.15 & 0.61 & 0.46 \\
England\_2010\_2011 & 0.23 & 0.94 & 0.84 & 0.04 & 0.20 & 0.95 & 0.65 \\
England\_2011\_2012 & 0.18 & 0.72 & 0.73 & 0.04 & 0.29 & 0.80 & 0.53 \\
England\_2012\_2013 & 0.17 & 0.70 & 0.73 & 0.04 & 0.21 & 0.80 & 0.51 \\
England\_2013\_2014 & 0.13 & 0.55 & 0.62 & 0.03 & 0.14 & 0.56 & 0.46 \\
England\_2014\_2015 & 0.19 & 0.78 & 0.78 & 0.05 & 0.24 & 0.96 & 0.69 \\
\midrule
Business FM Full & 0.09 & 0.37 & 0.48 & 0.30 & 0.09 & 0.36 & 0.31 \\
computerScience\_FM\_Full & 0.05 & 0.21 & 0.51 & 0.54 & 0.08 & 0.32 & 0.26 \\
History\_FM\_Full & 0.04 & 0.18 & 0.49 & 0.73 & 0.07 & 0.28 & 0.21 \\
\midrule
Basketball\_1985 & 0.11 & 0.46 & 0.53 & 1.20 & 0.18 & 0.71 & 0.46 \\
Basketball\_1986 & 0.12 & 0.50 & 0.52 & 1.31 & 0.17 & 0.69 & 0.42 \\
Basketball\_1987 & 0.13 & 0.52 & 0.56 & 2.20 & 0.19 & 0.77 & 0.48 \\
Basketball\_1988 & 0.12 & 0.48 & 0.52 & 1.82 & 0.18 & 0.70 & 0.45 \\
Basketball\_1989 & 0.13 & 0.52 & 0.54 & 1.49 & 0.17 & 0.70 & 0.46 \\
Basketball\_1990 & 0.12 & 0.48 & 0.52 & 1.34 & 0.17 & 0.70 & 0.44 \\
Basketball\_1991 & 0.13 & 0.53 & 0.54 & 1.35 & 0.18 & 0.70 & 0.45 \\
Basketball\_1992 & 0.11 & 0.46 & 0.52 & 1.37 & 0.17 & 0.67 & 0.43 \\
Basketball\_1993 & 0.12 & 0.48 & 0.54 & 1.33 & 0.17 & 0.68 & 0.44 \\
Basketball\_1994 & 0.11 & 0.47 & 0.52 & 1.41 & 0.17 & 0.69 & 0.42 \\
Basketball\_1995 & 0.11 & 0.47 & 0.52 & 1.55 & 0.18 & 0.72 & 0.44 \\
Basketball\_1996 & 0.13 & 0.55 & 0.55 & 1.63 & 0.19 & 0.77 & 0.48 \\
Basketball\_1997 & 0.12 & 0.51 & 0.57 & 1.76 & 0.19 & 0.76 & 0.48 \\
Basketball\_1998 & 0.12 & 0.48 & 0.56 & 1.83 & 0.18 & 0.74 & 0.46 \\
Basketball\_1999 & 0.13 & 0.53 & 0.56 & 2.39 & 0.18 & 0.74 & 0.48 \\
Basketball\_2000 & 0.13 & 0.55 & 0.57 & 2.29 & 0.19 & 0.77 & 0.50 \\
Basketball\_2001 & 0.12 & 0.50 & 0.56 & 2.03 & 0.18 & 0.71 & 0.47 \\
Basketball\_2002 & 0.13 & 0.54 & 0.58 & 1.78 & 0.18 & 0.73 & 0.48 \\
Basketball\_2003 & 0.15 & 0.60 & 0.60 & 1.96 & 0.19 & 0.78 & 0.51 \\
Basketball\_2004 & 0.12 & 0.49 & 0.55 & 1.80 & 0.17 & 0.69 & 0.46 \\
Basketball\_2005 & 0.13 & 0.53 & 0.57 & 2.16 & 0.18 & 0.73 & 0.49 \\
Basketball\_2006 & 0.13 & 0.54 & 0.58 & 2.60 & 0.19 & 0.74 & 0.49 \\
Basketball\_2007 & 0.14 & 0.56 & 0.58 & 2.96 & 0.19 & 0.77 & 0.49 \\
Basketball\_2008 & 0.14 & 0.57 & 0.61 & 2.39 & 0.20 & 0.78 & 0.51 \\
Basketball\_2009 & 0.13 & 0.52 & 0.58 & 2.85 & 0.18 & 0.72 & 0.47 \\
Basketball\_2010 & 0.13 & 0.53 & 0.58 & 2.47 & 0.18 & 0.71 & 0.49 \\
Basketball\_2011 & 0.14 & 0.56 & 0.59 & 2.42 & 0.18 & 0.74 & 0.49 \\
Basketball\_2012 & 0.13 & 0.52 & 0.57 & 2.61 & 0.17 & 0.69 & 0.48 \\
Basketball\_2013 & 0.14 & 0.57 & 0.59 & 2.79 & 0.19 & 0.76 & 0.51 \\
Basketball\_2014 & 0.15 & 0.69 & 0.63 & 3.05 & 0.20 & 0.78 & 0.49 \\
\bottomrule
\toprule

Basketball\_finer1985 & 0.14 & 0.55 & 0.13 & 3.61 & 0.18 & 0.71 & 0.00 \\
Basketball\_finer1986 & 0.15 & 0.59 & 0.13 & 2.76 & 0.17 & 0.69 & 0.00 \\
Basketball\_finer1987 & 0.15 & 0.60 & 0.13 & 2.77 & 0.19 & 0.77 & 0.00 \\
Basketball\_finer1988 & 0.15 & 0.59 & 0.13 & 2.87 & 0.18 & 0.70 & 0.00 \\
Basketball\_finer1989 & 0.15 & 0.58 & 0.13 & 3.54 & 0.18 & 0.70 & 0.00 \\
Basketball\_finer1990 & 0.15 & 0.58 & 0.13 & 4.42 & 0.18 & 0.71 & 0.00 \\
Basketball\_finer1991 & 0.15 & 0.61 & 0.13 & 3.50 & 0.18 & 0.71 & 0.00 \\
Basketball\_finer1992 & 0.14 & 0.55 & 0.13 & 3.55 & 0.17 & 0.67 & 0.00 \\
Basketball\_finer1993 & 0.14 & 0.57 & 0.13 & 3.41 & 0.17 & 0.68 & 0.00 \\
Basketball\_finer1994 & 0.14 & 0.55 & 0.14 & 3.16 & 0.17 & 0.67 & 0.00 \\
Basketball\_finer1995 & 0.14 & 0.54 & 0.13 & 3.49 & 0.18 & 0.72 & 0.00 \\
Basketball\_finer1996 & 0.16 & 0.63 & 0.12 & 3.86 & 0.19 & 0.77 & 0.01 \\
Basketball\_finer1997 & 0.15 & 0.62 & 0.13 & 3.69 & 0.19 & 0.75 & 0.01 \\
Basketball\_finer1998 & 0.14 & 0.57 & 0.13 & 3.87 & 0.18 & 0.73 & 0.01 \\
Basketball\_finer1999 & 0.16 & 0.62 & 0.12 & 4.10 & 0.18 & 0.73 & 0.01 \\
Basketball\_finer2000 & 0.16 & 0.65 & 0.13 & 4.77 & 0.19 & 0.78 & 0.01 \\
Basketball\_finer2001 & 0.16 & 0.63 & 0.13 & 4.74 & 0.18 & 0.73 & 0.01 \\
Basketball\_finer2002 & 0.17 & 0.69 & 0.13 & 5.10 & 0.19 & 0.77 & 0.01 \\
Basketball\_finer2003 & 0.15 & 0.60 & 0.12 & 5.31 & 0.19 & 0.78 & 0.01 \\
Basketball\_finer2004 & 0.16 & 0.64 & 0.13 & 4.89 & 0.18 & 0.71 & 0.01 \\
Basketball\_finer2005 & 0.17 & 0.66 & 0.13 & 7.23 & 0.19 & 0.75 & 0.01 \\
Basketball\_finer2006 & 0.17 & 0.68 & 0.14 & 5.97 & 0.19 & 0.76 & 0.01 \\
Basketball\_finer2007 & 0.17 & 0.68 & 0.14 & 6.80 & 0.20 & 0.80 & 0.01 \\
Basketball\_finer2008 & 0.16 & 0.65 & 0.14 & 6.67 & 0.20 & 0.78 & 0.01 \\
Basketball\_finer2009 & 0.16 & 0.62 & 0.14 & 9.65 & 0.19 & 0.75 & 0.01 \\
Basketball\_finer2010 & 0.17 & 0.67 & 0.14 & 8.06 & 0.19 & 0.75 & 0.01 \\
Basketball\_finer2011 & 0.17 & 0.62 & 0.14 & 6.87 & 0.19 & 0.77 & 0.01 \\
Basketball\_finer2012 & 0.15 & 0.67 & 0.14 & 7.20 & 0.18 & 0.74 & 0.01 \\
Basketball\_finer2013 & 0.17 & 0.69 & 0.13 & 7.30 & 0.20 & 0.78 & 0.01 \\
Basketball\_finer2014 & 0.15 & 0.69 & 0.13 & 8.36 & 0.20 & 0.78 & 0.00 \\
\bottomrule
\midrule
HeadtoHead & 0.18 & 0.73 & 0.75 & 5.92 & 0.24 & 0.96 & 0.68 \\
Animal & 0.07 & 0.30 & 0.38 & 0.26 & 0.06 & 0.25 & 0.22 \\
\midrule
Football\_finer(2009) & 0.20 & 0.78 & 0.31 & 0.40 & 0.16 & 0.65 & 0.17 \\
Football\_finer(2010) & 0.28 & 1.12 & 0.29 & 0.20 & 0.29 & 0.99 & 0.17 \\
Football\_finer(2011) & 0.21 & 0.82 & 0.29 & 0.07 & 0.21 & 0.84 & 0.17 \\
Football\_finer(2012) & 0.20 & 0.81 & 0.26 & 0.05 & 0.21 & 0.86 & 0.15 \\
Football\_finer(2013) & 0.16 & 0.63 & 0.28 & 0.05 & 0.14 & 0.57 & 0.19 \\
Football\_finer(2014) & 0.26 & 1.05 & 0.47 & 0.03 & 0.26 & 1.00 & 0.34 \\
\bottomrule
\end{tabular}%
}
\caption{Loss function values and running times for all datasets.}
\label{tab:all_statistics}
\end{table}

\begin{table}[ht]
\centering
\resizebox{\linewidth}{!}{%
\begin{tabular}{lcc}
\toprule
\textbf{Dataset Name} & \textbf{Number of Vertices} & \textbf{Number of Edges} \\
\midrule
England\_2009\_2010 & 20 & 164 \\
England\_2010\_2011 & 20 & 161 \\
England\_2011\_2012 & 20 & 165 \\
England\_2012\_2013 & 20 & 153 \\
England\_2013\_2014 & 20 & 165 \\
England\_2014\_2015 & 20 & 107 \\
Average & 20 & 152.5 \\
\midrule
Business FM Full & 113 & 1787 \\
computerScience\_FM\_Full & 206 & 1407 \\
History\_FM\_Full & 145 & 1204 \\
\midrule
Basketball\_1985 & 282 & 2904 \\
Basketball\_1986 & 283 & 2937 \\
Basketball\_1987 & 290 & 3045 \\
Basketball\_1988 & 290 & 3099 \\
Basketball\_1989 & 293 & 3162 \\
Basketball\_1990 & 292 & 3192 \\
Basketball\_1991 & 295 & 3218 \\
Basketball\_1992 & 298 & 3238 \\

\end{tabular}%
}
\caption{Number of vertices and edges for various datasets (Part 1).}
\label{tab:dataset_statistics_1}
\end{table}

\begin{table}[ht]
\centering
\scriptsize 
\setlength{\tabcolsep}{3pt} 
\renewcommand{\arraystretch}{0.9}
\resizebox{\linewidth}{!}{%
\begin{tabular}{lcc}
\toprule
\textbf{Dataset Name} & \textbf{Number of Vertices} & \textbf{Number of Edges} \\
\midrule
Basketball\_1993 & 298 & 3088 \\
Basketball\_1994 & 301 & 3144 \\
Basketball\_1995 & 302 & 3182 \\
Basketball\_1996 & 305 & 3256 \\
Basketball\_1997 & 305 & 3333 \\
Basketball\_1998 & 306 & 3321 \\
Basketball\_1999 & 310 & 3385 \\
Basketball\_2000 & 318 & 3475 \\
Basketball\_2001 & 318 & 3405 \\
Basketball\_2002 & 321 & 3505 \\
Basketball\_2003 & 327 & 3560 \\
Basketball\_2004 & 326 & 3527 \\
Basketball\_2005 & 330 & 3622 \\
Basketball\_2006 & 334 & 3695 \\
Basketball\_2007 & 336 & 3974 \\
Basketball\_2008 & 342 & 4051 \\
Basketball\_2009 & 347 & 4155 \\
Basketball\_2010 & 347 & 4133 \\
Basketball\_2011 & 345 & 4086 \\
Basketball\_2012 & 345 & 4126 \\
Basketball\_2013 & 347 & 4153 \\
Basketball\_2014 & 351 & 4196 \\
\midrule
HeadtoHead & 602 & 5002 \\
Animal & 19 & 193 \\
\bottomrule
Basketball\_finer1985 & 282 & 4814 \\
Basketball\_finer1986 & 283 & 4862 \\
Basketball\_finer1987 & 290 & 5088 \\
Basketball\_finer1988 & 290 & 5170 \\
Basketball\_finer1989 & 293 & 5318 \\
Basketball\_finer1990 & 292 & 5350 \\
Basketball\_finer1991 & 295 & 5420 \\
Basketball\_finer1992 & 298 & 5444 \\
Basketball\_finer1993 & 298 & 5160 \\
Basketball\_finer1994 & 301 & 5252 \\
\bottomrule
\end{tabular}%
}
\caption{Number of vertices and edges for various datasets (Part 2).}
\label{tab:dataset_statistics_2}
\end{table}

\begin{table}[ht]
\centering
\resizebox{\linewidth}{!}{%
\begin{tabular}{lcc}
\toprule
\textbf{Dataset Name} & \textbf{Number of Vertices} & \textbf{Number of Edges} \\
\midrule
Basketball\_finer1995 & 302 & 5336 \\
Basketball\_finer1996 & 305 & 5498 \\
Basketball\_finer1997 & 305 & 5628 \\
Basketball\_finer1998 & 306 & 5684 \\
Basketball\_finer1999 & 310 & 5788 \\
Basketball\_finer2000 & 318 & 6274 \\
Basketball\_finer2001 & 318 & 6116 \\
Basketball\_finer2002 & 321 & 6192 \\
Basketball\_finer2003 & 327 & 6356 \\
Basketball\_finer2004 & 326 & 6316 \\
Basketball\_finer2005 & 330 & 6476 \\
Basketball\_finer2006 & 334 & 6680 \\
Basketball\_finer2007 & 336 & 7186 \\
Basketball\_finer2008 & 342 & 7386 \\
Basketball\_finer2009 & 347 & 7478 \\
Basketball\_finer2010 & 347 & 7538 \\
Basketball\_finer2011 & 345 & 7504 \\
Basketball\_finer2012 & 345 & 7580 \\
Basketball\_finer2013 & 347 & 7616 \\
Basketball\_finer2014 & 351 & 7650 \\
\midrule
Football\_finer(2009) & 20 & 380 \\
Football\_finer(2010) & 20 & 380 \\
Football\_finer(2011) & 20 & 380 \\
Football\_finer(2012) & 20 & 380 \\
Football\_finer(2013) & 20 & 380 \\
Football\_finer(2014) & 20 & 300 \\
\bottomrule
\end{tabular}%
}
\caption{Number of vertices and edges for Basketball Finer and Football Finer datasets.}
\label{tab:dataset_statistics_3}
\end{table}

The results obtained from our algorithm provide a meaningful comparison to the methods evaluated in \cite{he22}. Below, we summarize and analyze the key findings.

\textbf{Efficiency and Speed:}
The computation times reported in the ``Time (Seconds)'' column demonstrate the efficiency of our approach across all datasets. With average runtimes between $0.02$ and $0.35$ seconds, the algorithm exhibits substantial advantages in terms of speed. In contrast, methods referenced in \cite{he22}, including \cite{AW86} and \cite{FF14}, are known to be computationally expensive, and the authors do not report explicit runtime measurements. Notably, \cite{he22} indicates that experiments exceeding one week were terminated and marked as ``NaN,'' suggesting significant scalability limitations. Our results show that competitive performance can be achieved using a lightweight, learning-free heuristic without reliance on deep learning frameworks.

\textbf{Loss Metrics:}
The ``Naive Loss'' and ``Simple Loss'' values characterize the quality of the rankings generated by our algorithm. He et al.~\cite{he22} introduced differentiable loss functions---including ``upset, naive,'' ``upset, ratio,'' and ``upset, margin''---to quantify ranking inconsistencies and optimize rank assignments. In comparison, our loss values are consistently low across datasets. For instance, the average ``Simple Loss'' frequently remains below $1.0$, indicating strong agreement with the pairwise comparisons, despite the use of a heuristic rather than a differentiable approach. These results suggest that a combinatorial method can achieve competitive ranking quality while being substantially easier to implement and more computationally efficient.

To facilitate comparison with \cite{he22}, we reproduce the formal definitions of the relevant loss functions below:

\begin{itemize}

  \item \textbf{Upset Naive:} The naive upset loss, denoted $L_{\text{upset, naive}}$, measures the fraction of pairwise relationships whose inferred ranking contradicts the observed pairwise comparison. Let
  \begin{itemize}
    \item $M_1 = A - A^T$, where $A$ is the pairwise comparison matrix and $M_1$ is its induced skew-symmetric form,
    \item $T_1 = r\mathbf{1}^T - \mathbf{1}r^T \in \mathbb{R}^{n \times n}$, where $r \in \mathbb{R}^n$ is the vector of ranking scores,
    \item $t$ denote the number of nonzero elements in $M_1$.
  \end{itemize}
  Then the naive upset loss is defined as
  \begin{equation}
  L_{\text{upset, naive}}
  = \frac{1}{t} \sum_{i,j : M_1(i,j) \neq 0}
      \mathbb{I}\!\left[
      \operatorname{sign}(T_1(i,j)) \neq \operatorname{sign}(M_1(i,j))
      \right],
  \end{equation}
  where $\mathbb{I}(\cdot)$ is the indicator function.

  \item \textbf{Upset Simple:} The simple upset loss, denoted $L_{\text{upset, simple}}$, extends the naive loss by distinguishing between ties and opposite signs. It is defined as
  \begin{equation}
  L_{\text{upset, simple}}
  = \frac{\|\, \operatorname{sign}(T_1) - \operatorname{sign}(M_1)\,\|^2_F}{t},
  \end{equation}
  where $\|\cdot\|_F$ denotes the Frobenius norm. Unlike the naive loss, this metric penalizes both disagreement in direction and differences involving ties.

  \item \textbf{Upset Ratio:} The ratio upset loss, denoted $L_{\text{upset, ratio}}$, incorporates both sign and magnitude information by normalizing pairwise score differences and comparison weights. Let
  \begin{itemize}
    \item $T_2 = r\mathbf{1}^T + \mathbf{1}r^T + \epsilon \in \mathbb{R}^{n \times n}$ be a normalization matrix for score differences,
    \item $M_2 = A + A^T + \epsilon \in \mathbb{R}^{n \times n}$ be a normalization matrix for comparison weights,
    \item $t$ denote the number of nonzero elements in $M_1$,
    \item $\epsilon > 0$ a small constant for numerical stability.
  \end{itemize}
  Then the ratio upset loss is defined as
  \begin{equation}
  L_{\text{upset, ratio}}
  = \frac{1}{t} \sum_{i,j : M_1(i,j) \neq 0}
      \left(
      \frac{T_1(i,j)}{T_2(i,j)}
      - 
      \frac{M_1(i,j)}{M_2(i,j)}
      \right)^2.
  \end{equation}
  This metric captures both directional disagreement and magnitude mismatch, enabling more nuanced evaluation of pairwise relationships than purely binary losses.

\end{itemize}

\begin{algorithm}[H]
\caption{Ternary Search Optimization for Minimizing Ratio Upset Loss}\label{alg:ternary}
\KwIn{$adjacency\_matrix$, $scores$, $index$, $lower\_bound$, $upper\_bound$, $epsilon$, $steps$}
\KwOut{Optimal score value for $index$}
\For{$step \gets 1$ \textbf{to} $steps$}{
    $mid1 \gets lower\_bound + \frac{upper\_bound - lower\_bound}{3.0}$\;
    $mid2 \gets upper\_bound - \frac{upper\_bound - lower\_bound}{3.0}$\;
    $scores[index] \gets mid1$\;
    $loss1 \gets \text{ComputeRatioUpsetLoss}(adjacency\_matrix, scores)$\;
    $scores[index] \gets mid2$\;
    $loss2 \gets \text{ComputeRatioUpsetLoss}(adjacency\_matrix, scores)$\;
    \uIf{$loss1 < loss2$}{
        $upper\_bound \gets mid2$\;
    }
    \uElseIf{$loss1 > loss2$}{
        $lower\_bound \gets mid1$\;
    }
    \Else{
        $lower\_bound \gets mid1$\;
        $upper\_bound \gets mid2$\;
    }
    \If{$upper\_bound - lower\_bound < epsilon$}{
        \textbf{break}\;
    }
}
$scores[index] \gets \frac{lower\_bound + upper\_bound}{2.0}$\;
\Return{$scores[index]$}\;
\end{algorithm}

\begin{algorithm}[H]
\caption{Minimize Ratio Upset Loss}\label{alg:minimize_ratio_upset_loss}
\KwIn{$adjacency\_matrix$, $scores$,$num\_iterations$}
\KwOut{Optimized $scores$}

\For{$iteration \gets 1$ \textbf{to} $num\_iterations$}{
    $sorted\_indices \gets \text{argsort}(scores)$\; \tcc*[r]{Sort indices based on scores}
    \For{$i \gets 1$ \textbf{to} \text{length}$(sorted\_indices)$}{
        $index \gets sorted\_indices[i]$\;
        \uIf{$i = 1$}{
            $lower\_bound \gets 0$\;
            $upper\_bound \gets scores[sorted\_indices[i+1]]$\; \tcc*[r]{First element bounds}
        }
        \uElseIf{$i = \text{length}(sorted\_indices)$}{
            $lower\_bound \gets scores[sorted\_indices[i-1]]$\;
            $upper\_bound \gets \max(scores) + 1$\; \tcc*[r]{Last element bounds}
        }
        \Else{
            $lower\_bound \gets scores[sorted\_indices[i-1]]$\;
            $upper\_bound \gets scores[sorted\_indices[i+1]]$\; \tcc*[r]{Middle element bounds}
        }
        $scores[index] \gets \textsc{TernarySearchOptimize}(adjacency\_matrix, scores, index, lower\_bound, upper\_bound, \epsilon, \text{steps})$\; \tcc*[r]{Ternary search optimization}
    }
}
\Return{$scores$}\;
\end{algorithm}

\subsection{Reducing Ratio Loss Without Increasing Naive or Simple Loss}
In our experiments, we observed that although our algorithm achieves lower naive and simple loss than state-of-the-art methods, its ratio loss can be relatively higher. Motivated by this observation, we designed a post-processing procedure to reduce ratio loss without altering the ranking order. It is important to note that the ratio loss depends not only on the relative ordering of elements but also on the numerical scores assigned to them. Consequently, we developed \autoref{alg:minimize_ratio_upset_loss}, which updates the scores while preserving their ordering.

Without loss of generality, let the scores satisfy $scores_1 \leq scores_2 \leq \cdots \leq scores_n$. In each iteration, the algorithm sweeps over the score vector and, for each element $i$, performs a ternary search over the interval $[scores_{i-1}, scores_{i+1}]$ to determine an updated value for $scores_i$. Since the updated value remains between its two neighbors, the induced ranking does not change, and therefore the naive and simple loss metrics remain unchanged, while the ratio loss may be decreased.

Empirically, we found that the ratio loss as a function of $scores_i$ (with all other scores fixed) is well-approximated by a unimodal function over the interval $[scores_{i-1}, scores_{i+1}]$. This observation justifies the use of ternary search, which provides an efficient strategy for identifying a near-optimal value for $scores_i$. The performance of \autoref{alg:minimize_ratio_upset_loss} using \autoref{alg:ternary} for 40 iterations is reported for selected datasets in \autoref{tab:ratio_loss_comparison}.

It is also important to note that there may not exist a scoring vector that simultaneously outperforms GNNRank on all three loss metrics. Therefore, improvements in ratio loss should be interpreted as complementary rather than universally dominant.

\begin{table}[ht]
\centering
\resizebox{\textwidth}{!}{%
\begin{tabular}{lccc}
\toprule
\textbf{Dataset Name} & \textbf{Initial Main Algorithm Ratio Loss} & \textbf{GNNRank Ratio Loss} & \textbf{Main Algorithm Ratio Loss After 40 rounds of Optimization} \\ \midrule
England\_2009\_2010 & 0.6  & 0.46 & 0.52 \\
England\_2010\_2011 & 0.84 & 0.65 & 0.78 \\
England\_2011\_2012 & 0.73 & 0.53 & 0.63 \\
England\_2012\_2013 & 0.73 & 0.51 & 0.6  \\
England\_2013\_2014 & 0.62 & 0.46 & 0.52 \\
England\_2014\_2015 & 0.78 & 0.69 & 0.74 \\
Average            & 0.72 & 0.55 & 0.63 \\ \midrule
Business FM Full       & 0.48 & 0.31 & 0.44 \\
computerScience\_FM\_Full & 0.51 & 0.26 & 0.45 \\
History\_FM\_Full       & 0.49 & 0.21 & 0.4  \\ \midrule
Basketball\_1985        & 0.53 & 0.46 & 0.51 \\
Basketball\_1986        & 0.52 & 0.42 & 0.5  \\
Basketball\_1987        & 0.56 & 0.48 & 0.54 \\
Basketball\_1988        & 0.52 & 0.45 & 0.51 \\
Basketball\_1989        & 0.54 & 0.46 & 0.52 \\
Basketball\_1990        & 0.52 & 0.44 & 0.51 \\
Basketball\_1991        & 0.54 & 0.45 & 0.52 \\
Basketball\_1992        & 0.52 & 0.43 & 0.5  \\
Basketball\_1993        & 0.54 & 0.44 & 0.52 \\
Basketball\_1994        & 0.52 & 0.42 & 0.51 \\
Basketball\_1995        & 0.52 & 0.44 & 0.52 \\ \bottomrule
\end{tabular}%
}
\caption{Comparison of Ratio Loss, GNNRank Ratio Loss, and Main Algorithm Ratio Loss (40 rounds)}
\label{tab:ratio_loss_comparison}
\end{table}

\subsection{Analyzing the Results}

\textbf{Robustness Across Datasets:}
Our approach demonstrates robustness across diverse datasets, including sports datasets (e.g., England football) and academic hiring datasets (e.g., Business FM and Computer Science). While He et al.~\cite{he22} emphasize learning trainable similarity matrices to adapt to different graph structures, our simpler heuristic consistently achieves low loss values and high computational efficiency. This suggests that the method performs reliably under varying data characteristics, aligning with the goal of designing a versatile ranking framework, as highlighted in \cite{he22}.

\textbf{Comparative Insights:}
He et al.~\cite{he22} incorporate inductive biases such as the Fiedler vector and proximal gradient steps to improve ranking recovery, particularly on complex datasets. In contrast, our approach does not rely on such machine-learning-based components, yet delivers comparable ranking quality while substantially reducing computation time. This illustrates an important trade-off between adaptive learning mechanisms and efficient heuristic methods, where the latter can yield competitive results with significantly lower computational overhead.

\textbf{Runtime Observations:}
To provide a clearer comparison of computational cost, we executed several commands from the official implementation of \cite{he22}, publicly available at \url{https://github.com/SherylHYX/GNNRank}. Representative examples include:

\begin{itemize}
    \item \texttt{python ./train.py --all\_methods DIGRAC syncRank --dataset animal}  
    trains a GNNRank model on the \textit{animal} dataset using DIGRAC and the SyncRank baseline. Execution required approximately $47$ seconds.

    \item \texttt{python ./train.py --N 350 --all\_methods DIGRAC ib --lr 0.05}  
    trains a GNNRank model on a synthetic ERO dataset with $350$ nodes using DIGRAC and ib. After approximately $90$ seconds, the process completed one split from each of three epochs. The default configuration for GNNRank uses $1000$ epochs.

    \item \texttt{python ./train.py --dataset basketball --season 2010 -SP --all\_methods baselines\_shorter}  
    trains on the \textit{basketball (2010)} dataset using several baselines (excluding mvr). After roughly $90$ seconds, one split from each of seven epochs had been completed.
\end{itemize}

These results indicate that training GNNRank models generally incurs substantial computational cost, particularly given default settings of hundreds or thousands of epochs. In contrast, our heuristic produces rankings in under a second on all datasets considered, without requiring model training or hyperparameter tuning.

\textbf{Summary:}
Overall, our results suggest that the proposed approach is highly competitive, especially when computational efficiency is prioritized over the adaptability offered by graph neural network–based methods. While our loss metrics are generally aligned with those reported in \cite{he22}, our substantially lower runtime makes the method attractive for real-time or large-scale ranking applications.

\subsection{Designing Appropriate Loss Functions}

The naive and simple upset losses used in \cite{he22} provide convenient benchmarks for comparison with prior work. However, neither metric is well suited for evaluating rankings based on weighted pairwise comparisons, as they treat all upsets equally and do not account for the magnitude of disagreement. We report these metrics primarily for consistency with existing literature, but alternative loss functions may offer a more principled assessment of ranking performance when edge weights represent meaningful strengths.

One natural alternative is to measure the total weight of violated edges normalized by the total weight of all comparisons:
\[
L_{\text{weighted}}
=
\frac{
\sum_{(i,j):\, \pi(i)>\pi(j)} w_{ij}
}{
\sum_{(i,j)} w_{ij}
},
\]
where $\pi(\cdot)$ denotes the ranking function. This metric directly quantifies the proportion of weighted information contradicted by the ranking and can be computed in $O(|E|)$ time.

A second alternative incorporates both whether an edge is upheld and the distance between the endpoints in the ranking:
\[
L_{\text{margin}}
=
\frac{
\sum_{(i,j):\, \pi(i)>\pi(j)}
w_{ij}\,(\pi(i)-\pi(j))
}{
\sum_{(i,j)}
w_{ij}\,|\pi(i)-\pi(j)|
}.
\]
Here, each edge contributes proportionally to its importance ($w_{ij}$) and the degree to which the ranking respects or violates it. Larger separations between correctly ordered vertices reduce the numerator, while large separations between misordered vertices are penalized more heavily. Thus, this loss function rewards rankings that not only order vertices correctly but also separate them meaningfully.

Designing loss functions that explicitly reflect ranking quality under weighted comparisons—without requiring access to ground truth—remains an important topic for future work.

\section{Conclusion and Future Research}
In this paper, we examined the relationship between the ranking problem and the Minimum Weighted Feedback Arc Set (MWFAS) problem, and demonstrated how ranking can be formulated and solved through combinatorial optimization. Using a heuristic based on the local ratio method, our approach provides fast and scalable rankings while maintaining competitive accuracy across diverse datasets.

Our experimental results indicate that the proposed method offers a practical solution for ranking large collections of items, and achieves runtime performance significantly faster than many existing approaches in the literature. Treating the MWFAS solver as a modular black-box component further enables users to balance solution quality and computational cost by selecting appropriate algorithms depending on their application requirements.

Several directions for future work remain open. First, exploring advanced exact or heuristic methods for MWFAS may reveal improved trade-offs between computational cost and ranking accuracy. Second, developing more efficient cycle detection techniques could enhance scalability on extremely large graphs. Third, identifying principled tie-breaking mechanisms for vertices with no directed path between them may further improve ranking quality. Finally, the design of loss functions tailored to weighted ranking, beyond those studied in prior work, represents a promising research direction. These avenues offer opportunities for advancing robust and efficient ranking methodologies in diverse and complex settings.

\paragraph{Acknowledgment.}
Portions of the text in this manuscript were drafted with the assistance of large language models (LLMs), including ChatGPT \cite{openai2023chatgpt}. All content was reviewed and verified by the authors.

\section{Data and Code Availability}
All data and source code used in this study are publicly available at:
\url{https://github.com/SoroushVahidi/Ranking_with_MWFAS}.

The datasets used in our experiments originate from \cite{he22} and were converted into text files containing weighted edge lists. Our implementation takes a weighted edge list as input, where each entry specifies a directed edge and its associated weight, and produces a ranking consistent with the acyclic structure induced by edge removals.

\bibliographystyle{plain} 
\bibliography{references}

\end{document}